\documentclass[aps,prl,twocolumn,showpacs,superscriptaddress]{revtex4}  

\usepackage{graphicx}  
\usepackage{dcolumn}   
\usepackage{bm}        
\usepackage{amssymb}   
\usepackage{multirow}

\newcommand{\be}{\begin{equation}}
\newcommand{\ee}{\end{equation}}
\newcommand{\ba}{\begin{eqnarray}}
\newcommand{\ea}{\end{eqnarray}}
\newcommand{\ban}{\begin{eqnarray*}}
\newcommand{\ean}{\end{eqnarray*}}
\newcommand{\p}{\paragraph{}}
\newcommand{\fig}[1]{ (Fig.~\ref{#1})}

\newcommand{\ket}[1]{\mbox{$ | #1 \rangle $}}
\newcommand{\bra}[1]{\mbox{$ \langle #1 | $}}

\def\opone{\leavevmode\hbox{\small1\kern-3.8pt\normalsize1}}

\newcommand{\one}{\leavevmode\hbox{\small1\normalsize\kern-.33em1}}

\begin{document}

\title{Simple synchronization of independent picosecond photon sources for quantum communication experiments}


\author{O. Landry \footnote[2]{email:olivier.landry@unige.ch}}
\author{J.A.W. van Houwelingen}
\affiliation{Group of Applied Physics, University of
Geneva, rue de l'Ecole-de-M\'{e}decine 20,1211, Gen\`{e}ve,
Switzerland\\}
\author{P. Aboussouan}
\affiliation{Laboratoire de Physique de la Mati\`ere Condens\'ee, CNRS UMR 6622,
Universit\'e de Nice Sophia-Antipolis, Parc Valrose, 06108 Nice Cedex 2, France}
\author{A. Beveratos}
\affiliation{Laboratoire de Photonique et Nanostructures LPN-CNRS UPR-20, Route de Nozay, 91460
Marcoussis, France}
\author{S. Tanzilli}
\affiliation{Laboratoire de Physique de la Mati\`ere Condens\'ee, CNRS UMR 6622,
Universit\'e de Nice Sophia-Antipolis, Parc Valrose, 06108 Nice Cedex 2, France}
\author{H. Zbinden}
\author{N. Gisin}
\affiliation{Group of Applied Physics, University of
Geneva, rue de l'Ecole-de-M\'{e}decine 20,1211, Gen\`{e}ve,
Switzerland\\}

\date{\today}

\begin{abstract}
An essential requirement for the future of quantum communication
is the capability to create undistinguishable photons at distant
locations. The main challenge is to control the various sources of timing jitter in order to conserve temporal undistinguishability.
 We report on a system that enables a
large tolerance against such jitter by using synchronized sources
of photons with coherence lengths on the order of centimeters.
\end{abstract}

\pacs{03.67.Hk,42.81.-i,42.65.Re} \maketitle \p

\section{1. INTRODUCTION}

\begin{figure}[]
\includegraphics[scale=0.25]{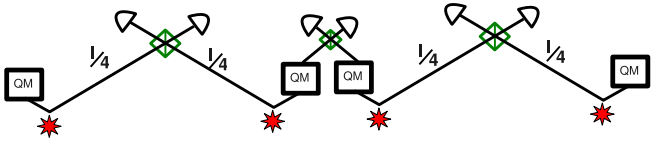}
\caption{Graphic representation of a quantum repeater of total length $l$. The squares
represent quantum memories and the stars photon pair sources. At
midpoint between the sources a beamsplitter performs a simple BSA for which synchronization is
required. \label{quantumrepeater}}
\end{figure}

Quantum repeaters (QR) \cite{Briegel1998, Duan2001,Sangouard2009} are devices that use sources of entanglement\ \cite{Bouwmeester1997, Boschi1998,
Furusawa1998, Kim2001, Lombardi2002, Marcikic2003,Pomarico2009}, quantum memories\ \cite{Kraus2006,Afzelius2009,Zhao2009,Hosseini2009} and partial Bell state analyzers\ \cite{Weinfurter1994,Brendel1999,Houwelingen2006} (BSA) to perform
cascaded entanglement swappings over a large distance between photons that never interacted \ \fig{quantumrepeater}.

One particular difficulty is the synchronization of independent sources of entangled pairs of photons. Partial BSA 
functions through perfect undistinguishability of two photons at a beamsplitter. If the two
photons arrive at different times at the beamsplitter the BSA does not
succeed, and therefore neither does the entanglement swapping.
Several methods have been conceived to ensure temporal
undistinguishability. 

Previous experiments have achieved this using a femtosecond laser as a common pump\ \cite{Pan1998, Riedmatten2005}. However this method is not applicable to realistic implementations of a quantum repeater which must have independent nodes. The simple  duplication of femtosecond lasers on different locations is difficult to achieve as the sources must be synchronized to a better precision than their pulse lengths. For a typical Ti-Sapphire pulse length of 200\ fs (with coherence length of the same order), this corresponds to 40\ $\mu$m. Path length fluctuation of underground fibers, which is on the order of 10$^{-5}\ \text{K}^{-1}$, or 1\ cm/(km$\cdot$K), would destroy indistinguishability at the BSA even in the presence of perfect sychronization\ \cite{CRC,Minar2008}. Such experiments have nevertheless been performed in the laboratory using either cavities with a common element\ \cite{Yang2006} or actively controlled cavities with fast electronics\ \cite{Kaltenbaek2006, Kaltenbaek2009}.

\begin{figure}[]
\includegraphics[scale=1]{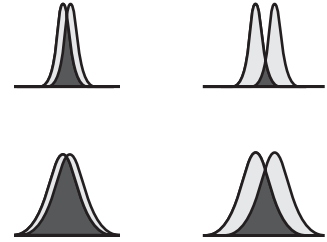}
\caption{Effects of a displacement on the overlap between two
wavepackets. The top two pictures show a clear reduction of the
overlapping area (black) of two identical but displaced wavepackets
when their displacement is increased. The bottom two pictures show a
less important reduction when the wavepackets are
wider.\label{jitter}}
\end{figure}

Spectral filtering of the photon pairs increases their coherence length and relaxes the synchronization requirements, as shown in Fig.~\ref{jitter}. In turn, this requires a narrow pump to conserve energy-time entanglement. If pulsed operation is not necessary, narrowly filtered continuous sources can be used and post-selected entanglement swapping resistant to path length fluctuations has been demonstrated\ \cite{Halder2008,Yang2009}.

In this article we present independent synchronised sources of pulsed picosecond length entangled photons. One laser is an electrically
pumped pulsed diode laser which can be triggered externally. This is a simpler and less expensive solution than the usually used modelocked lasers. The second laser is a modelocked Ti-Sapphire in the picosecond regime.

We demonstrate first that our lasers are synchronized at a distance of 2.2\ km with a timing jitter of less than  60\ ps. Then we create down-converted photons using these lasers and additional filtering, and we perform a Hong-Ou-Mandel (HOM) dip to measure the degree of undistinguishability of these photons. We demonstrate that long distance synchronization does not affect undistinguishability.

\section{2. Diode Laser Source}

\begin{figure}[]
\includegraphics[scale=0.5]{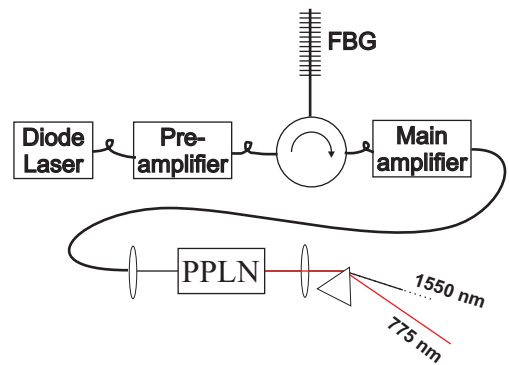}
\caption{Schematic of the diode based pulsed picosecond laser. The
light from a pulsed diode laser is amplified in several stages.
After this it is frequency doubled and the remaining pump photons are
spatially filtered using a dispersion prism.\label{sourcepq}}
\end{figure}

It is important to have simple, easily deployable entanglement sources. With this in
mind we developed a source based on a pulsed laser diode, instead
of the commonly used modelocked lasers \fig{sourcepq}. The diode laser (PicoQuant LDH-P-C-W-1550 with PDL-800-B driver) produces pulses of light
with a FWHM of about 20\ ps at a wavelength of 1550\ nm with 20\ $\mu$W mean power with 12.3\ mW peak power at a repetition rate of 76\ MHz. A higher output power is possible but this
increases the pulse length. In order to amplify the signal
three Er-doped fiber amplifiers are used. First, the light passes
two pre-amplifiers (one home-made, the other model EAD-60-C from IPG). The amplifiers are operated in the linear regime to minimize spontaneous amplified emission (ASE). The output is filtered with a 1\ nm spectral width Fiber Bragg Grating (FBG) in order to
remove the residual ASE, after which it has a mean power of 1\ mW. Afterwards the light is once more amplified by 25.4\ dB using the main amplifier (Keopsys high-power amplifier). The
resulting pulses have a measured mean power of 350\ mW.
These pulses are then frequency-doubled by a periodically poled lithium niobate (PPLN) crystal which generates 2\ mW mean power or 1.0$\cdot$10$^8$ photons/pulse with 1.2\ W peak power of the desired light at 775\ nm
with a spectral width of 0.4\ nm.

\section{3. Source synchronization}

In a quantum repeater setup such as that of Fig.~\ref{quantumrepeater}, many sources will need to be synchronized such that the produced photons are in the same temporal mode at the beamsplitters. We demonstrate here synchronization of the diode laser in slave mode triggered by a signal emitted by a MIRA modelocked laser (Coherent) generating 5 picosecond long pulses with a wavelength of\ 775nm and a bandwidth of 1.1\ nm.

We use two different links between the master and the slave. As a first step, we direct a small part of the light emitted from the modelocked laser to a high-speed InGaAs photodetector (Thorlabs, DET01CFC). The output signal is directly used to externally trigger the diode laser.

\begin{figure}
\includegraphics[scale=0.5]{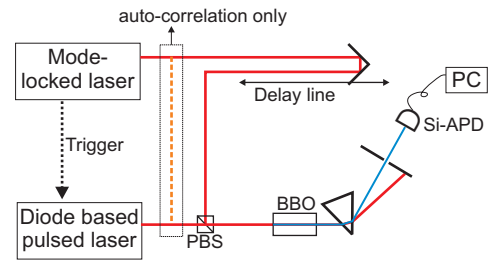}
\caption{Setup for correlation measurements. For
cross-correlations two laser pulses from different sources are
combined at a polarization beamsplitter (PBS). For autocorrelation measurements these pulses
originate from the same laser. Frequency doubling is only possible
if both pulses pass through the crystal at the same
time.\label{correlation}}
\end{figure}

\begin{figure}[!b]
\includegraphics[width=8cm]{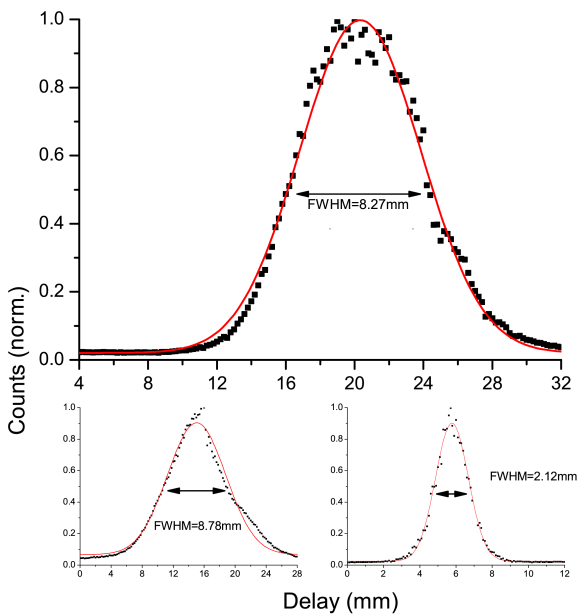}
\caption{Results of the cross- and auto-correlation measurements.
\textit{top}: Crosscorrelation with a gaussian fit FWHM=8.27\ mm, or 27.6\ ps.
\textit{bottom-left}: Autocorrelation of the diode laser based source,
corresponds to a pulse length of 20.7\ ps per pulse. \textit{bottom-right}:
Autocorrelation of the modelocked source, corresponds to a pulse length
of 5.0\ ps per pulse. \label{resultcorrelation}}
\end{figure}

Before we can test the synchronization we must perform an auto-correlation measurement \fig{correlation} to establish the pulse lengths of both pump lasers. We separate the beams in two parts, one of which passes through an optical delay line. Both beams are then recombined at a PBS, such that the time delay between the horizontal and vertical polarizations of the output mode of the PBS is controlled by the delay. This output mode is sent through a beta-barium borate (BBO) crystal cut for type II sum frequency generation
of two 775\ nm photons of perpendicular polarizations to a single 387.5\ nm photon. Such generation is only possible if the input modes arrive at the same time at the PBS. The generated blue light is
separated from the other wavelengths with a prism and measured
using a silicon photon counting module  (idQuantique ID100). The delay is scanned while recording the
count rate. The resulting autocorrelation function \fig{resultcorrelation} shows the modelocked laser pulses have a sech$^2$ shape with a FWHM of 5.0\ ps while the diode laser pulses have roughly gaussian shape with FWHM of 20.7\ ps.

We then perform a cross-correlation measurement by sending the output of each laser on a different input mode of the PBS. Again, blue light is only generated if pulses from both lasers arrive at the same time. Jitter between the arrival times will show up as a broadening of the cross-correlation function, with total width equal to
$\tau_{cc}=\sqrt{\tau_{1}^2+\tau_{2}^2+\tau_{sync}^2}$ where $\tau_{cc}, \tau_{1},
\tau_{2}$ and $\tau_{sync}$ are the pulse lengths FWHM for the cross-correlation function,
first laser pulse shape, second laser pulse shape and the synchronization jitter distribution's FWHM
respectively. We measure $\tau_{cc}$=27.6\ ps, for a resulting $\tau_{sync}$=18\ ps.

\begin{figure}[hbt]
\includegraphics[width=8cm]{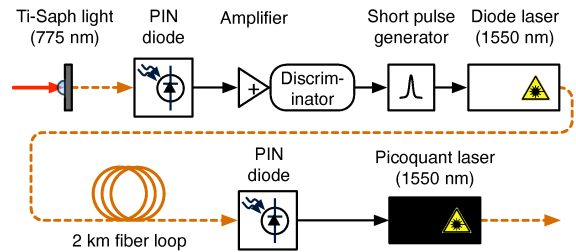}
\caption{Media Converter layout. \label{mediaconverter}}
\end{figure}

\begin{figure}[hbt]
\includegraphics[width=8cm]{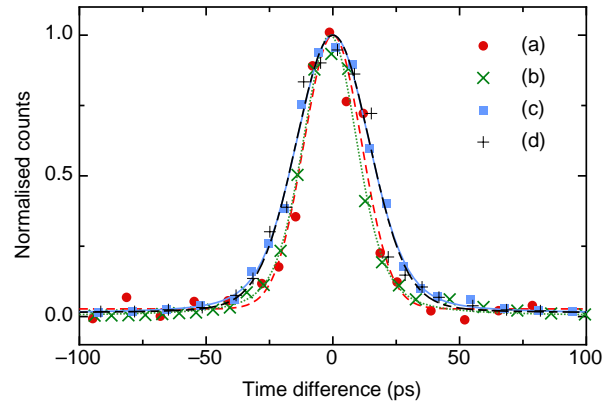}
\caption{Cross-correlation function using different synchronization systems. (a) Direct photodetector signal. (b) Amplified photodetector signal. (c) Media Converter with 50\ cm fiber. (d) Media Converter with 2.2\ km fiber.  \label{mediaconvertercc}}
\end{figure}

Electrical impulses are impractical over long distances, as jitter increases with losses in the cable. A more versatile method is to use a media converter to send an optical synchronization signal over large distances. We have built such a converter, shown in Fig.\ \ref{mediaconverter}. The output signal from the InGaAs high-speed photodetector is first amplified by 7\ dB before being converted to an ECL signal. Adding the amplifier alone does not add measurable jitter. This signal triggers a standard telecom laser (Bookham) which sends corresponding pulses through a fiber, which are then detected by another InGaAs high-speed photodetector. The output from this diode is then used to trigger the diode laser. Cross-correlations measurements were performed with fiber lengths of 50\ cm and 2.2\ km \fig{mediaconvertercc}. The widths of the cross-correlation were 34.6\ ps irrespective of the fiber length, for a jitter induced by the media converter of 21\ ps and a total jitter for the full synchronization line of 27\ ps.

\begin{figure}[]
\includegraphics[width=6cm, angle=270]{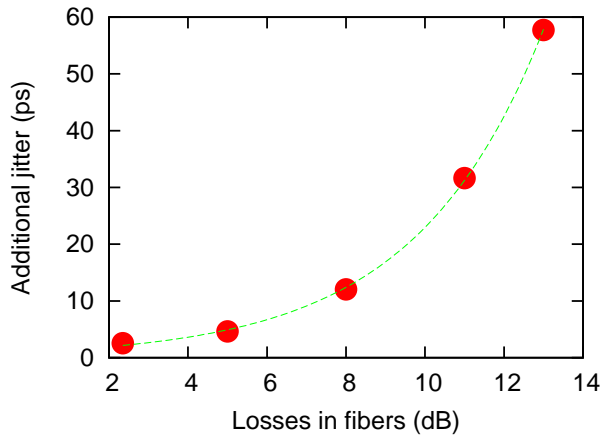}
\caption{Additional jitter as a function of fiber loss. \label{triglength}}
\end{figure}

With longer fibers, higher jitter can be caused by a lower signal-to-noise ratio on the InGaAs photodetector due to losses. Such additional jitter is shown in Fig. \ref{triglength} as measured directly on a 6 GHz oscilloscope through the electric signal jitter using an optical variable attenuator. The exact values can depend strongly on both the pulse shape and the quality of the photodetector in the media converter.

In the field, the jitter will also be increased by path length fluctuations. Studies have shown that, over the course of a day, we can expect a fluctuation on the order of $10^{-5}$ for commercially installed underground fiber\ \cite{Minar2008}, i.e. cm-length fluctuations for every km of fiber. For example, if we want to see fluctuations of less than the pulse length of the diode laser, we could only tolerate 0.4\ km. For larger distances, active fibre length stabilization may be required. For 36\ km, for example, fibre length would have to be measured and stabilized every 6 minutes, compared to every second if we were to use a femtosecond source, which is a clear advantage.

\section{4. Production of undistinguishable photon pairs}

We can use these pulsed lasers and spontaneous
parametric downconversion (SPDC) to create undistinguishable photon pairs useful for quantum communication. We can then check undistinguishability by performing a HOM dip \cite{Hong1987a}.

\begin{figure}[]
\includegraphics[scale=0.5]{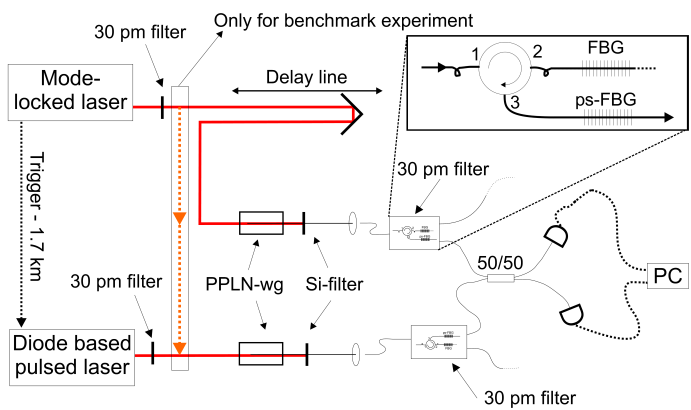}
\caption{Experimental setup used to measure a HOM dip with
independent sources. The photons combine at the 50/50 beamsplitter
and will bunch depending on the delay line.\label{HOMdip}}
\end{figure}

The setup for the HOM dip \fig{HOMdip} consists of the same two synchronized lasers as used for the
cross-correlation measurements. The light from these lasers is
sent into PPLN waveguides in order to produce pairs of photons
using SPDC. The generated paired photons are then collected into optical 
fibers, spectrally filtered and separated.

In any case the coherence length and therefore the pulse length of the created photon pairs must
be larger than the length of the pump pulse that generated them
\cite{Zukowski1995}. Our pump pulses were filtered using a 30\ pm bulk bandpass filters (Layertec), enlarging them to a 29\ ps coherence time, slightly above their original durations, to make them Fourier-transform limited and ensure time-bin entanglement.

To separate signal and idler photons out of each photon pair generator, we use filtering
stages made of standard fiber optic components (AOS Gmbh), i.e., the combination of a fiber Bragg
grating (FBG) reflecting the desired wavelength of 1548 nm within a large
bandwidth, a circulator and a phase shifted FBG as depicted in the inset of
Fig. \ref{HOMdip}. The latter FBG features a 30 pm large transmission peak at 1548 nm from its
otherwise broadband reflexion. Two such filtering stages, one at the output of each
photon pair source, are employed and fine-tuned so as to match each other using thermal
expansion. The signal photons spectrally filtered that way show a resulting coherence
time of 117\ ps.
Note that the idler photons, around the wavelength of 1552 nm, are not used in this
experiment.

These indistinguishable signal photons are then sent to a beamsplitter. The
variable delay is scanned and coincidence count rates recorded.

\section{5. Expected visibility}

The visibility of the resulting HOM dip will be a measure of the undistinguishability of the photons produced. The effect of jitter on the visibility is given by eq.\ \ref{vjitter}, calculated in the appendix, and depends on $\text{r}_{\text{j}}$, the FWHM of the distribution of the timing jitter, in units of the coherence length.

\ba%
\overline{V}=&\frac{1}{\sqrt{1+\frac{1}{2}r_j^2}}
\label{vjitter}
\ea%

\begin{figure}[]
\includegraphics[angle=270,width=0.4\textwidth]{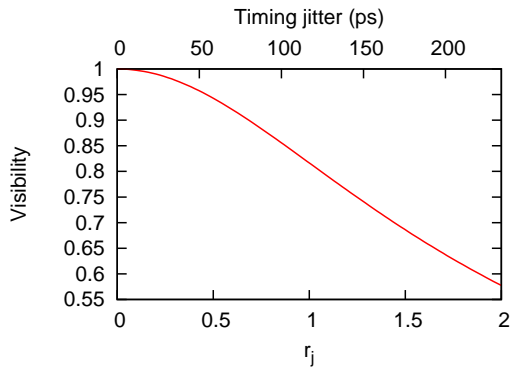}
\caption{Average visibility as a function of the dimensionless ratio $\text{r}_{\text{j}}$ and as a function of the timing jitter for the specific case of a 117\ ps coherent pulse length. \label{Vvsjitter}}
\end{figure}

This result shows a clear decrease of the
visibility as a result of time-of-arrival jitter\ \fig{Vvsjitter}. It also shows
that a small non-zero value of $r_j$ can be tolerated without a dramatic loss of
visibility.
In our case, considering the pump pulse length of 29\ ps and the synchronization jitter of 27\ ps, the total timing jitter is 50\ ps. With a coherence length of 117\ ps, we expect a visibility of 96\%.
The visibility of a HOM dip directly indicates the maximal
visibility that can be obtained in an entanglement swapping
experiment. 

If we consider Fourier-transform limited gaussian pulses, the 
coherence length $l_c$ is given by\ \cite{Saleh1991}
\ba%
l_c=0.44\cdot\frac{\lambda_0^2}{\Delta\lambda},
\ea%
where $\lambda_0$ is the central wavelength and $\Delta\lambda$ is the
spectral FWHM. In order to increase the coherence length and thus
effectively reduce $r_j$, a small $\Delta\lambda$ is necessary.

\section{6. HOM dip}

\begin{figure}[]
\begin{center}
\includegraphics[angle=-90,width=7cm]{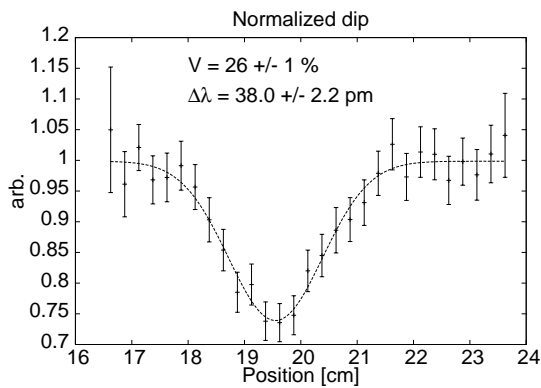}
\caption{Measured HOM dip using two independent and separated
sources.\label{MIRAPICO}}
\end{center}
\end{figure}

The count rates are recorded using a time-to-digital converter (TDC) both for true coincidences ($\sim1.4\ s^{-1}$) and for photons which did not arrive at the same time at the beamsplitter, allowing a measure of the noise due to unsynchronized dark counts. The noise is substracted and the resulting dip \fig{MIRAPICO} has a net visibility of $V=0.26\pm0.011$ and a width corresponding to $126\pm8$\ ps. 
Note that the maximum visibility that can be found is
0.33 because the sources are
probabilistic\ \cite{Riedmatten2003}, so that we reach 78\% of the expected value.

The loss of visibility cannot be due to inaccurate synchronization. Considering the photon's coherence length and the measured jitter, the width of the dip should have been 173\ ps. To fully account for the reduced visibility, a jitter of 128\ ps would be needed, which would correspond to a dip width of 209\ ps.

Additionally, in order to confirm that the cause of the low visibility was not
jitter, we performed a HOM dip using only the modelocked laser as the common pump of both SPDC sources, with no synchronization. In
this setup there is no synchronization jitter from the pump photons. The resulting visibility and dip widths were the same.

These two results combined
lead to the conclusion that the loss of visibility is not caused
by jitter and our media converter synchronization works as intended.

Other possible reasons for the low visibility have been investigated. The probability of creating a pair per pulse has been set to approximately 10\% in both sources, so the production of extra pairs should increase the visibility by at most 0.9\ \%. Spectral undistinguishability has been measured with careful characterization of the filters. The difference between their spectral width leads to a decrease in visibility of at most 1\% and their central wavelength has been shown to be stable over many weeks. 
Pair production statistics has been measured, following the theory presented in \cite{Riedmatten2004b}, and the results show that stimulated emission occurs as expected. A different result would have led to an increase of visibility.
Walk-off in the waveguide due to group velocity dispersion has been calculated to be 7.3\ ps, entailing a visibility reduction of 0.2\ \%. Polarization is adjusted using polarization controllers and a polarizer, for an estimated visibility reduction of 3\ \%.
The total visibility, taking into account all these sources of distinguishability in the worst scenario, should have been at least 0.306, far from the experimental result.

We note that our results are comparable to those of other groups working with waveguides, for example Laiho et al. working with periodically poled potassium titanyl phosphate (PPKTP) waveguides and a common source also achieved a 78\% maximal fidelity\ \cite{Laiho2009}. Such visibilities are sufficient to demonstrate nonlocality\ \cite{Branciard2009}. On the other hand, some of us achieved a 99\% net visibility using separated  PPLN waveguides pumped by a common modelocked laser\ \cite{Aboussouan2009}.


\section{7. Conclusions}

In conclusion, we have shown that it is possible to build pulsed
photon sources with picosecond pulse lengths using different
pumps. It is possible to synchronize two sources to such an
extent that the jitter and path length fluctuations are not a
cause of loss of visibility. Such sources are required to build quantum repeaters with multiple locations for the photon
sources. 

\section{Acknowledgements}

The authors thank Claudio Barreiro, Thomas Ganz, Jean-Daniel Gauthier, Marc P. de Micheli, Bruno Sanguinetti, Christoph Simon and Sorin Tascu for their
support and input. This work was supported by the NCCR Quantum Photonics
(NCCR QP), research instrument of the Swiss National
Science Foundation (SNSF), and by the European Research Council through the Qore Advanced Grant.

\bibliography{GAP}

\appendix
\section{APPENDIX: HOM DIP WITH JITTER}

Consider a beamsplitter for which the input modes are labelled $a^{\dagger}$ and $b^{\dagger}$, while the output modes are $c^{\dagger}$ and $d^{\dagger}$ such that the action of the beamsplitter is

\ba
a^{\dagger}\to \frac{1}{\sqrt{2}} (i c^{\dagger} + d^{\dagger}) \\
b^{\dagger}\to  \frac{1}{\sqrt{2}} (c^{\dagger} + i d^{\dagger} ). 
\ea

If photons come in from either side separated by a time delay $\tau$, the corresponding state and its projection in time will be

\ba
\ket{\Psi_i} =& a_0^{\dagger} b_{\tau}^{\dagger}\ket{0} \\
=&\int \psi(\omega_a) \psi(\omega_b)e^{i\omega_b \tau} \\
& \cdot \ket{\omega_a} \ket{\omega_b} d\omega_a d\omega_b \\
\text{s.t.}\hspace{2mm} \bra{t_a}\bra{t_b}\ket{\Psi_i} =& \psi(t_a) \psi(t_b-\tau) 
\ea

where $\psi(t)$ is a gaussian wavepacket and $\psi(\omega)$ its Fourier transform. The output state will be

\ba
\ket{\Psi_o} =& \frac{1}{2} [i c_0^{\dagger}c_{\tau}^{\dagger}+i d_0^{\dagger}d_{\tau}^{\dagger}\\
& +c_0^{\dagger}d_{\tau}^{\dagger}-c_{\tau}^{\dagger}d_{0}^{\dagger}]\ket{0} \\
=&\frac{1}{2} [i \int \psi^2(\omega_a) e^{i \omega_a \tau}\ket{\omega_a} d\omega_a \ket{0} \\
&+i \int \psi^2(\omega_b) e^{i \omega_b \tau}\ket{\omega_b} d\omega_b \ket{0} \\
& +\int \psi(\omega_a)\psi(\omega_b) (e^{i \omega_a \tau}-e^{i \omega_b \tau})\\
& \cdot \ket{\omega_a} \ket{\omega_b}d\omega_ad\omega_b ].
\ea

For $\tau$ much larger than the wavepacket length the last two terms are completely independent; for $\tau=0$ they cancel each other and the photons bunch. The amplitude corresponding to photons exiting the beamsplitter from different outputs $\ket{\Psi_{do}}$ is

\ba
\bra{t_a} \bra{t_b} \ket{\Psi_{do}} = \psi(t_a) \psi(t_b-\tau)-\psi(t_a-\tau) \psi(t_b),
\ea

such that the total probability of photons not bunching is

\ba
P=\int\int [\psi(t_a) \psi(t_b-\tau)-\psi(t_a-\tau) \psi(t_b)]^2 dt_a dt_b.
\ea

If we define $\psi(t)=\frac{1}{\sqrt{\sigma\sqrt{2\pi}}} e^{\frac{t^2}{4\sigma^2}}$, such that $2\sqrt{2\ln{2}}\sigma$ is the pulse width at FWHM, we get

\ba
P(\tau)=2\cdot(1-e^{-\frac{\tau^2}{4\sigma^2}}).
\ea

The resulting visibility is

\ba
V(\tau)=&\frac{P(\infty)-P(\tau)}{P(\infty)} \\
=& e^{-\frac{\tau^2}{4\sigma^2}}.
\ea

If we rewrite this with the dimensionless ratio $\Delta=\tau/\sigma$

\ba
V(\Delta)=e^{-\frac{\Delta^2}{4}},
\ea

and if we define a probability distribution for this ratio

\ba
\rho(\Delta) = \frac{1}{\sigma_{\Delta}\sqrt{2\pi}} e^{-\frac{\Delta^2}{2\sigma_{\Delta}^2}},
\ea

we can find the average visibility

\ba
\overline{V}=&\int \rho(\Delta)V(\Delta) d\Delta\\
=&\frac{1}{\sqrt{1+\frac{1}{2}\sigma_{\Delta}^2}}.
\ea

We also perform the same calculations using FWHMs instead of variances. We use the variables

\ba
\Delta'=&\frac{2\sqrt{2\ln{2}}\tau}{\sigma} \\
r_j =& \frac{\sigma_{\Delta}}{2\sqrt{2\ln{2}}}.
\ea

The final result takes the same form in new units:

\ba
\overline{V}=&\frac{1}{\sqrt{1+\frac{1}{2}r_j^2}}.
\ea

\end{document}